# Unraveling the different regimes arisen during plasma ammonia synthesis on mesoporous silica SBA-15 through plasma diagnostics

Sophia Gershman[1], Henry Fetsch[1], Fnu Gorky[2], Maria L. Carreon[3]*

[1] Princeton Plasma Physics Laboratory, 100 Stellarator Road, Princeton, New Jersey 08540

[2] Chemical and Biological Engineering Department, South Dakota School of Mines & Technology, 501 E Saint Joseph St, Rapid City, South Dakota 57701, USA

[3] Mechanical Engineering Department, University of Massachusetts Lowell, One University Avenue, Lowell Massachusetts 01854-5043, USA

*Corresponding author: maria_carreon@uml.edu

**Abstract**: Herein we demonstrate that the performance of mesoporous silica SBA-15 and SBA-15-Ag impregnated during plasma ammonia synthesis depend on the plasma conditions. At high power the mesoporous silica SBA-15 without Ag produces the largest amount of ammonia observed in our experiments, but the addition of Ag provides a minor benefit at lower powers. Plasma conditions were analyzed using optical emission spectroscopy using $N_2$, $N_2^+$ and NH molecular bands and $H_\alpha$ lines. The analysis of optical emission, specifically Ha lines and $N_2$ molecular bands data shows that SBA-15 has higher electron density, and higher $N_2$ vibrational temperature. Stronger $N_2^+$ emission relative to the neutral $N_2$ provides additional evidence for higher electron density in the SBA-15 packed bed reactors. The addition of Ag resulted in a lower concentration of atomic hydrogen. Where Ag acts as a hydrogen sink facilitating surface reactions with nitrogen. The presence NH in the spectrum may indicate a higher concentration of H in the SBA-15 reactor. From the materials point of view SBA-15 is a more robust catalyst with a commendable surface area retention after plasma exposure due to the lack of local heating generated when a metal is in the structure.

**Introduction**

Ammonia is among the most essential synthetic chemicals. It sustains global food production due to its use in the production of fertilizers. Consequently, ammonia access impacts crop yields and the food market.[1] To improve ammonia availability in remote areas, there are several ongoing efforts to develop small-scale, non-centralized ammonia synthesis[2,3] which would cost less than transporting it from centralized locations while also benefiting the local economy. Ammonia decentralization strategies could also provide long-term carbon-free fuel[4,5] due to its potential as carbon free vector for hydrogen.[6] In fact, ammonia can be competitive with gasoline, Compressed Natural Gas (CNG), Liquid Petroleum Gas (LPG) and methanol in gravimetric, volumetric and energetic costs.[7] Hence, a possible development of an infrastructure for ammonia storage, handling, and decentralized *synthesis*, can lead to a future so-called "ammonia economy."[8,9,10] Currently, the Haber-Bosch (HB) process dominates the industrial ammonia production. This process requires high temperatures and pressures, ~500 ºC and 500 bar, and large centralized production plants to be economic viable.[11,12] Consequently, the current industrial ammonia synthesis structure hampers the access to affordable fertilizers for farms in remote areas.[13] Its energy requirements make the HB the most energy-consuming process in the chemical industry. In addition to fertilizers, ammonia is an essential chemical employed as a precursor for paints and cleaning agents, which explains its high global production (~249.4 million tons) that requires ~1-2% of the world's energy, 2-3% of the world's natural gas output, and results in the release of over 300 million metric tons of $CO_2$[14,15] (~1.5 % of all greenhouse gas emissions globally[2]). In this respect, the development of simplified alternatives to HB at milder conditions and compatible with intermittent electric power (e.g., from renewable energy sources) is a critical step toward small-scale, decentralized ammonia production.

Recently, the development of plasma catalysis, the synergistic effect of plasma with a heterogeneous catalyst, has been driven by the prospect of pairing certain plasma reactors with inexpensive renewable electricity sources[16] hence, reducing the carbon footprint. The cost-effective implementation of solar and wind sources observed during the past decade motivates the search for sustainable alternatives for plasma catalytic synthesis of ammonia. However, the potential implementation of non-thermal plasma technology will be feasible until there is a competitive plasma catalytic system in terms of material selection and reactor design.

From the materials standpoint, oxides are robust and efficient low-cost materials for catalytic applications. Oxides besides their low cost, offer textural, compositional, and morphological properties that can be tailored for diverse targeted catalytic reactions.[17] From our recent experimental and DFT calculations,[18] one of our central hypotheses is that an optimal catalyst for plasma-assisted ammonia synthesis is one that delays the recombination of *adsorbed* hydrogen radicals (H*) into molecular hydrogen ($H_2$), allowing them to bind instead to *adsorbed* nitrogen plasma generated species ($N_2^+$) to form NH*. Therefore, our interest in oxides stems from the possibility of facilitating hydrogen dissolution[19] at mild reaction conditions.[20, 21, 22]

These oxide catalysts such as silicas and aluminas have been employed before for plasma catalytic ammonia synthesis.[23, 24, 25] However, there is still much to understand when using ordered porous oxides in plasma-based processes. Interestingly it has been reported that the ammonia production could be effectively enhanced by introducing oxide-type catalysts with low electronegativity[26] (approx. 3.38 for the case of silica).[27] Moreover, most of the existing plasma research on porous materials, concludes on the benefits of the porosity, including the formation of microdischarges[28, 29, 30, 31, 32] and short-lived species in the inter-particle spaces.[28, 29] A feature exploited in this work through the use of a doughnut shape mesoporous silica SBA-15 which provide such spacing. Apart from these advantages, porous materials with ordered pores, in the mesoporous range (2-50 nm) such as silica, are appealing materials for plasma catalytic ammonia synthesis due to their: 1) $SiO_2$ lower electrical resistivity that can lead to more stable and uniform plasma discharges[33]; 2) readiness to dissolve hydrogen[19]; 3) weakly bonding with nitrogen[34]; 4) high thermal and chemical stability in the presence of water and some hydrocarbons, which are typical impurities in natural gas wells (hydrogen for ammonia production is typically obtained from steam reforming of methane[35]), 5) high surface areas that can lead to improved mass transfer between guest-product species avoiding mass transfer limitations, and 6) Depending on the porous material a rich variety of chemical wall compositions is possible. Therefore, molecule adsorption and transport properties can be fine-tuned by the chemistry of the porous material. In addition, the desorption processes of the products, in this case ammonia, play a key role. Ammonia binding energies tend to be higher on oxide surfaces than on metal surfaces; typically ranging from -32 kJ/mol to -547 kJ/mol more than 2 times the values for metals (-13.50 kJ/mol to -96.48 kJ/mol). Hence, ammonia desorption is easier from metallic surfaces compared to oxide surfaces. Furthermore, in porous oxides like SBA-15 ammonia adsorption can be high due to the large available surface area. A feature that can be exploited in the future to protect ammonia from further decomposition by collision with energetic particles from plasma while working on optimized pulses. On the other hand, oxides have the desirable characteristic as a catalyst for plasma synthesis, of weakly bonding nitrogen since metals have higher nitrogen binding energies than oxides such as silica.

The benefits of having a hierarchically ordered mesoporous structure in heterogeneous catalysis is well documented[36, 37] and encouraged our interest in such materials. Specifically, surfactant-directed self-assembly approach is an effective simple strategy for the development of mesoporous oxides with desirable structural, compositional, and morphological properties.[38, 39, 40]

Recently, our group demonstrated the importance of the pore size by comparing microporous materials, including metal organic frameworks,[41] zeolites[42, 43] and mesoporous silica SBA-15[44] acting as effective catalysts for the plasma catalytic synthesis of ammonia. Specifically, plasma synthesis assisted by SBA-15, a mesoporous silica with unimodal pore size of ~ 7 nm, resulted in an ammonia synthesis rate per gram of catalyst at least 3 times higher than that of the non-porous and fumed silica when employing a hydrogen rich gas feed.

From the reactor design, dielectric barrier packed bed reactors offer the benefits of operating at atmospheric pressure, limiting the bulk heating in the reactor by limiting the discharge current, the possibility of operating from a renewable energy power source, and access to a variety of discharge regimes that may be beneficial for the plasma catalyst interaction.[45] Dielectric barrier discharges can take diffuse or microdischarge mode and transition to filamentary form with an application of longer pulses and higher voltage.[46, 47, 48] In a packed bed reactor, the discharge can start in a bulk mode with microdischarges forming from the electrode to the beads and between the dielectric beads due to field enhancement by polarization and surface charging. As mentioned above the microdischarges are instrumental in producing the active species such as vibrationally excited nitrogen and H*. As the applied voltage increases, surface ionization waves become more likely and the discharge transitions to a surface mode where the filaments form along the surface of the beads.[49, 50, 51] The surface mode of the discharge may be highly beneficial for catalytic function of the packing materials. Recent modeling results show the importance of vibrational excitation of nitrogen molecules in predicting the observed yield of ammonia in plasma synthesis suggesting that lower plasma powers may be more efficient at utilizing both the plasma energy and the catalysts.[52] These studies motivate the exploration of plasma conditions as well as catalytic materials.

The goal of this work was to address two hypotheses, one, that the best performing metal catalysts in plasma catalysis differ in their properties from the ones used in thermal catalysis. Metals were chosen based on our previously reported simulation results and their ability to dissolve hydrogen as well as their ability to attract nitrogen.[18, 53] The second hypothesis explored is that mesoporous oxides may be sufficient for the plasma catalytic process without the aid of a metal catalyst therefore presenting an alternative to metal catalysis.

**Experimental Section**

*Catalyst synthesis*

<u>SBA-15 Synthesis</u>

Tetraethyl orthosilicate (TEOS, Aldrich 98%), Pluronic P123 (P123, BASF), deionized water, and hydrochloric acid (Macron Fine Chemicals), N, N-dimethylformamide (DMF, 99.8%, Sigma Aldrich) were employed for the synthesis of SBA-15. The SBA-15 catalyst was synthesized based on the synthetic protocol reported elsewhere.[54] Briefly, 2.0 g of P123 was combined with 30 g of water, 15 g DMF and stirred at 40 °C. Next 30 g of 4 M HCl and 4.4 g of TEOS were added. The solution was stirred at 40 °C for 20 h, and then the resultant gel was transferred to a Teflon liner and heated at 100 °C for 24 h in an oven. The product was removed from the oven and cooled before vacuum filtering. The final product was dried at room temperature overnight and was washed with deionized water. Finally, the structure directing agent P123 was removed by calcination in air at 500 °C for 6 h.

<u>SBA-15-Ag synthesis</u>

The employed impregnation method was modified from elsewhere.[55] Briefly, 2 g of SBA-15 was mixed into an aqueous solution of $AgNO_3$ (0.1 M, 36.5 mL deionized water). The solution was sonicated for 5 minutes then stirred constantly at room temperature. Subsequently, the pH of the solution was adjusted to 8.5 by dropwise addition of 3% aqueous $NH_3$ solution. After achieving the required pH, the solution was again sonicated for 5 min then constantly stirred at room temperature for 3 hours. Further, the sample was filtered, rinsed three times with deionized water, and dried at 100 °C overnight.

*Calcination and reduction*

2 g of sample was loaded into a vacuum chamber. The sample was calcined at 450 °C for 1 hour in air (0.5 torr, 10°C min$^{-1}$, ramp), purged with nitrogen for 30 min at 450 °C (0.5 torr), reduced at 450 °C for 1 h in hydrogen plasma (0.5 torr, 80 watts), purged again with nitrogen at 450 °C (0.5 torr), and cooled under nitrogen flow to room temperature (0.5 torr, 10 °C min-1 ramp). The calculated theoretical silver load on SBA-15 is 10 % wt.

**Catalyst Characterization**

Surface area, pore size and pore volume were analyzed using the micromeritics Gemini II—2375 BET (Brunauer-Emmett-Teller) surface area analyzer. Fresh and spent were degassed at 300 °C for 5 h before analysis. Nitrogen adsorption/desorption isotherms and the corresponding pore size distribution of SBA-15 agreed with the ones reported in our previous work on this silica.[44] These isotherms correspond to Type IV with H1-type hysteresis, which is characteristic of mesoporous phases.

The morphology of the catalyst samples was analyzed using transmission electron microscopy (TEM) in a Tecnai F20 (FEI, Inc.) microscope operating at 200 kV. TEM specimens were prepared by dispersing fine powders of catalyst and reference samples on standard holey carbon-on-copper 300 mesh TEM grids. Scanning transmission electron microscopy (STEM) images were collected from these specimens using a high annular angle dark filed (HAADF) detector (E.A. Fischione Instruments, Inc.) and a sub-nanometer electron probe. The use of STEM-HAADF imaging provided a two-fold advantage. First, it enabled a strong Z-contrast, and allowed to clearly distinguish between the oxide support grains composed of light elements (O, Si) and the metal catalyst nanoparticles, characterized by much higher atomic number (47 for Ag as opposed to 8 and 14 for O, and Si, respectively). Second, by using STEM imaging, the total electron-beam exposure of the samples was significantly reduced as compared to imaging in the regular TEM mode. Thus, the effects such as radiation damage of beam sensitive oxide grains or beam-induced agglomeration of catalyst nanoparticles were significantly suppressed.

*Experiment setup*

Two DBD reactor designs were used in this study i.e., with and without a powdered catalyst. Referred here, respectively, as packed-bed and plug reactors.[42] The reactor setup was similar to the one we reported previously.[56] Each DBD reactor consisted of a metal electrode (2.4 mm diameter) extended 30 mm into the quartz tube (4 mm I.D. and 6.35 mm O.D.) and a tinned copper mesh ground electrode (23.5 mm length) fitted around the outside of the tube. The quartz tube served as the dielectric barrier. Assuming a straight and well centered inner electrode, there was 0.8 mm clearance between the metal electrode and the inner walls of the quartz tube that remained free for the working gas to flow in the plug flow reactors or it was filled with the mesoporous catalyst in the packed bed reactors (**Figure 1**). The working gas consisted of a mixture of $N_2$ and $H_2$, at the mass flow ratios of 1:1 and 1:3, and a constant volumetric flow rate of 25 sccm (mass flow controllers, MKS 946) for all experiments and reactors.

Power was supplied to the reactor by a PVM500-1000 power supply (Information Unlimited, LLC). The power supply was operated at a 40% duty cycle and pulse-on repetition rate of 100 Hz. The ac signal frequency was adjusted to achieve resonance of the output secondary transformer of the power supply and the plasma reactor as a parallel load. The resonance frequency was 44+/-2 kHz. The <2 kHz frequency adjustments made to remain in resonance under slightly changing operational conditions such as heating of the system.

Ammonia production was monitored using Fourier transform infrared absorption spectroscopy (FTIR-AS); electrical measurements provided the operational conditions, and plasma parameters were determined using optical emission spectroscopy (**Figure 1**).

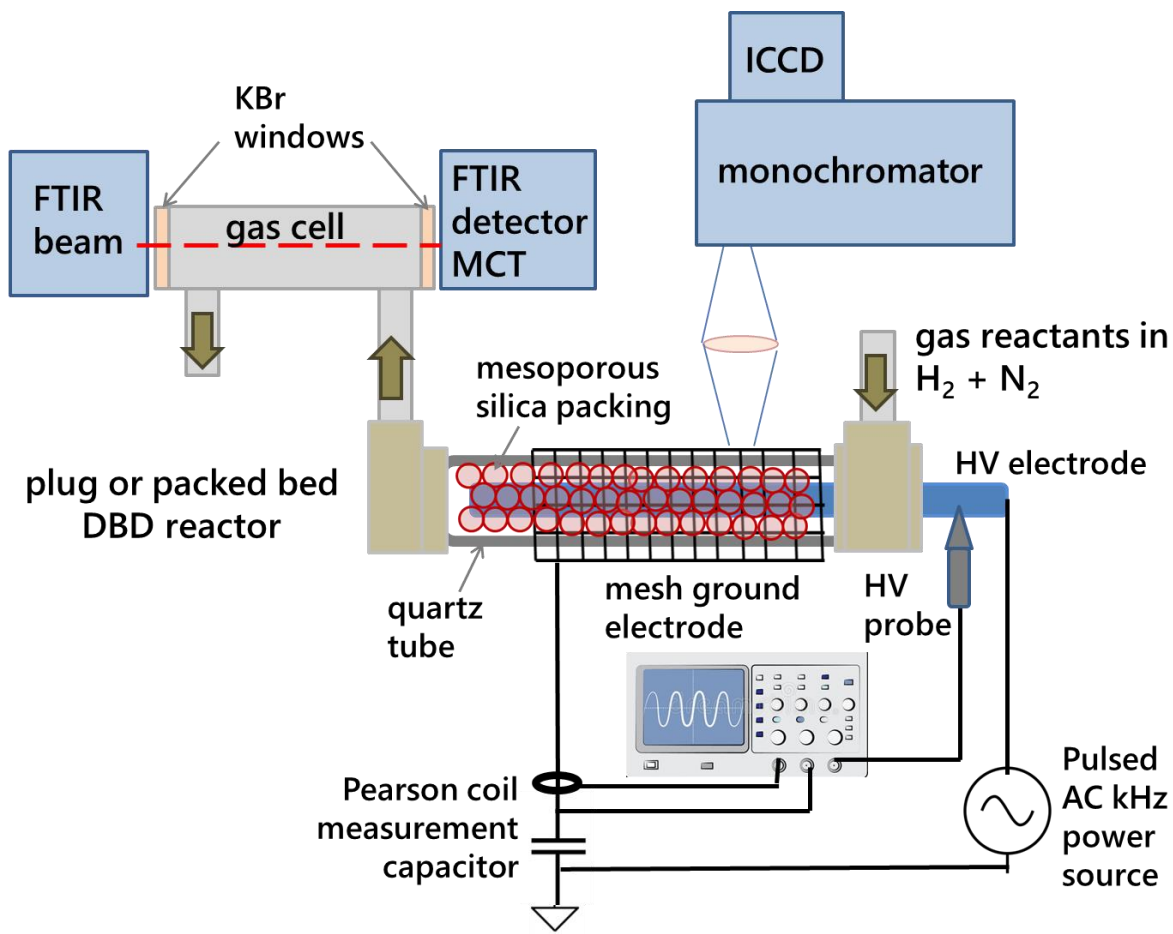

**Figure 1.** Schematic of the experimental setup including the reactor, electrical diagnostics, optical emission spectroscopy, and FTIR gas output measurement. ⬇⬆ - gas flow of reactants in and products out of the reactor and through the gas cell for FTIR analysis.

## Results and Discussion

*Catalyst Stability*

The employed materials were characterized by Scanning Electron Microscopy (SEM). They displayed the characteristic doughnut like shape morphology for this type of SBA-15 samples (See Supporting Information). EDX analysis was performed on the SBA-15-Ag to confirm the presence of Ag (**Figure S1**). To gain a better understanding of the morphology we performed Transmission Electron Microscopy (TEM) to the fresh and spent samples. **Figure 2** shows the TEM images for SBA-15. The images confirm the doughnut like morphology. Moreover, this analysis allowed us to not only observe a reduction of the pore size for the spent samples. But to witness a change in the doughnut morphology after plasma exposure. When looking at the spent samples it is evident that the doughnut shape has evolved to a distorted or squeezed doughnut (see **Figure S2**). In terms of the pore size for the fresh SBA-15 sample the pore is in average 7.79 nm while the spent sample showed an average pore size of 7.75 nm (0.04nm reduction). (See **Table 1**, and **Figure S3-5** showing SBA-15-Ag TEM image & pore size distributions). Similar reduction was observed for the SBA-15-Ag sample. In this case the pore size for the fresh sample went from 8.16nm to 7.14nm (1.02 nm reduction). However, not only the pore size reduction can be accounted for the surface area reduction but the distorted doughnut shapes after plasma exposure.

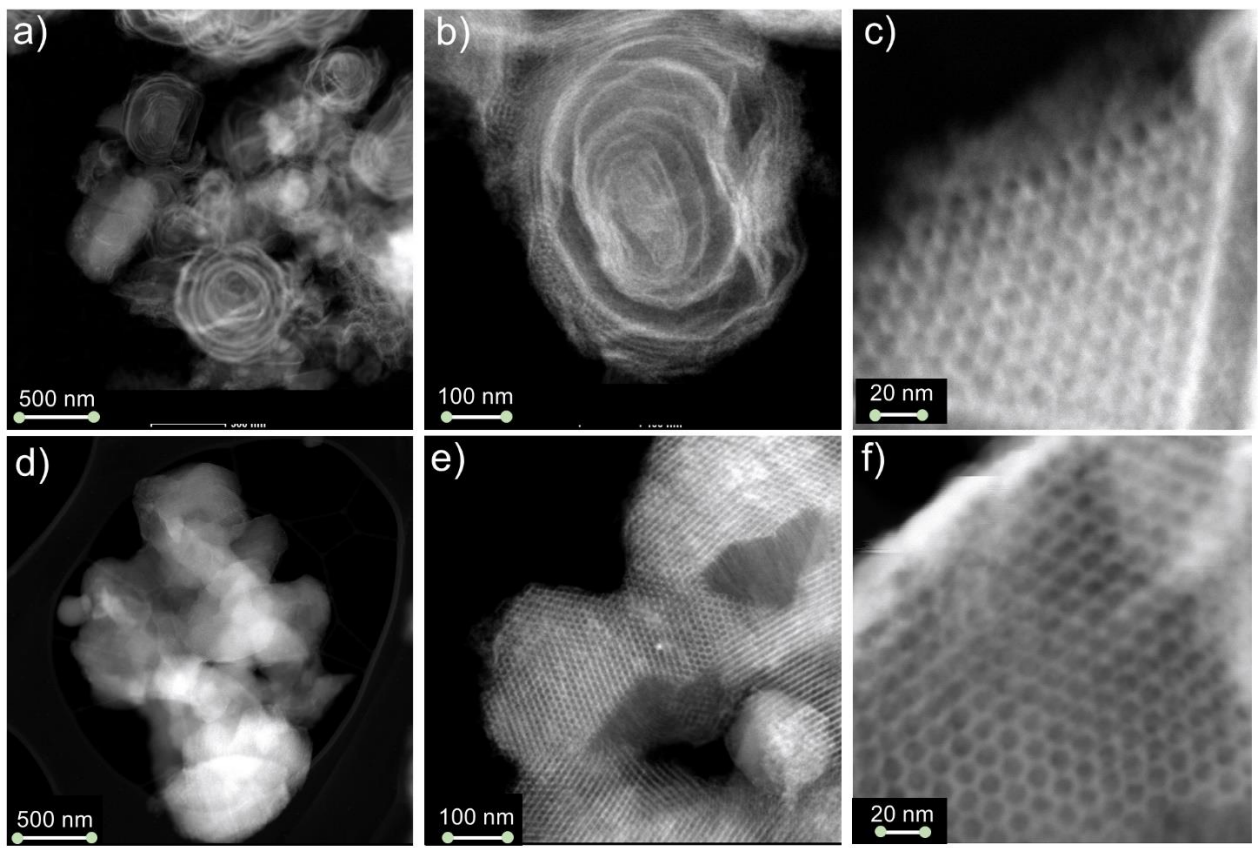

**Figure 2.** Transmission Electron Microscopy (TEM) images for fresh SBA-15 at a) 500nm, b) 100nm and c) 20nm and for spent SBA-15 (after plasma exposure) at d) 500nm, e) 100nm and f) 20nm.

The BET data also shows a considerably higher reduction on the surface area for the SBA-15-Ag sample compared to the sample with no Ag on it (see **Table 1**). This higher lost on the surface area for the SBA-15-Ag can be attributed to the formation of hot spots in the metal nanoparticles that act as heating concentrators leading to a more structurally affected material. The SBA-15-Ag showed nanoparticle clustering after plasma exposure. Specifically, the particle size increased after plasma exposure increased from 3.6 nm to 6.5 nm. (see **Figure S6**) Our group has reported this kind of behavior for a silica-nickel system for plasma ammonia synthesis.[57]

**Table 1.** Morphological characterization of the catalysts employed in this study.

| Catalyst | | BET Surface Area | Pore Volume | Pore Size (BET) | Pore Size (TEM) | Surface area reduction |
|---|---|---|---|---|---|---|
| *Units* | | *(m² g⁻¹)* | *(cm³ g⁻¹)* | (nm) | (nm) | % |
| SBA-15 | Fresh | 884.23 | 1.26 | 7.1 | 7.79 ± 0.82 | 41.88 |
| | Spent | 513.9 | 0.91 | 8.0 | 7.75 ± 1.08 | |
| SBA-15-Ag | Fresh | 963.04 | 2.13 | 8.1 | 8.16 ± 0.65 | 60.40 |
| | Spent | 381.32 | 0.72 | 7.7 | 7.14 ± 0.87 | |

*Electrical Measurements*

Voltage on the high voltage electrode, current to ground, and the charge on the measurement capacitor (**Figure 1**) were monitored during all experiments. A P6015A (Tektronix, Inc.) high-voltage probe used for recording the applied voltage, can compensate for capacitances 7 – 49 pF, suitable for the DBD reactor here. Current to ground was measured by a Pearson current monitor model 2877, Pearson Electronics. The voltage, $\Delta V$ across the monitoring capacitor was measured by a standard grounded oscilloscope probe with no attenuation or a 10-fold attenuation as needed, and the charge was calculated as usual, $Q=C\Delta V$. The value of the monitoring capacitor, C=10 nF, was chosen to be much greater than the system capacitance of tens of pF, so that it doesn't interfere with the electrical operation of the DBD discharge.

Power consumption was calculated from the electrical data using Lissajous figures[ref.58, 59] The charge on the measurement capacitor was plotted against the applied voltage for one full period, yielding a closed curve as seen in **Figure 8**. The area enclosed by this curve is the energy dissipated in the reactor during one cycle. The power is then given by the frequency *f* and the duty cycle *D*:

$$P = f \cdot D \cdot \oint q dV,$$

where the integral is taken over one ac cycle and the value is averaged over several cycles for each measurement. During reactor operation, oscilloscope traces of voltage, current, and charge were recorded on two time scales, 5 µs, to monitor the ac applied signal, and 2.5 ms to determine the duty cycle, D, of the applied voltage. Resonance frequency, f, was determined by adjusting the ac frequency at a constant output setting on the power source while monitoring the applied voltage trace. Since the power source is used in a parallel resonance configuration, at resonance, the applied voltage is a maximum with minimal distortion of the waveform.

***Monitoring Ammonia Production***

*Measurement and Interpretation of the FTIR-AS*

FTIR-AS was used to analyze the reactor products. We used a single-beam FTIR-6800 spectrometer with a liquid nitrogen-cooled MCT detector (Jasco) at 1 cm$^{-1}$ resolution. The measurements were conducted using a 10 cm optical path gas cell (Pike Technologies) with 25 mm diameter KBr windows. The gas cell was located inside the FTIR sample compartment, 50 cm downstream from the DBD reactor.

Before FTIR measurements, the reactor and gas cell were flushed using the working gas mixture for 30 minutes. A background spectrum was taken while flushing, immediately before the plasma was lit in each trial. Once the reactor was turned on and brought to full power, FTIR spectra were recorded at approximately three-minute intervals for the duration of each trial of about fifteen minutes. Each spectrum was composed of 128 separate scans averaged internally by the instrument.

To calculate ammonia concentration, we simulated the absorption spectrum of ammonia using the single-molecule line-by-line parameters from the HITRAN database.[60] The line intensities were adjusted to account for the temperature-dependent population in the lower state of each transition. Gas temperature was set as an input parameter. Several temperatures were tested, and the best fit was found 325K (see temperature dependence section in the Supporting Information file.). Small pressure shifts were applied to each line's location, and then each line was broadened to incorporate physical and instrument effects. First each line was broadened into a Lorentzian; in this system at atmospheric pressure, the pressure broadening contribution dominated over the Gaussian doppler broadening[47]. The resulting spectrum was then smoothed to account for the instrument resolution. A cosine apodization function was used on the FTIR spectrometer, and so we convolved the physical spectrum with the Fourier transform of the apodization function to simulate the spectrum measured at the detector.

The simulated spectra are in good agreement with those that we obtained experimentally. A representative spectrum, obtained from a SBA-15 reactor with silver, is shown in **Figure 3** along with the simulated ammonia spectrum. A scaling factor was applied to the simulated spectrum to match the height of the measured spectrum. This factor is equal to the number of ammonia molecules per unit area in the beam path. Because the optical path length inside the gas cell was fixed at 10 cm, and the gas flow was mostly uniform along the length, the scaling factor could be used to determine concentration.

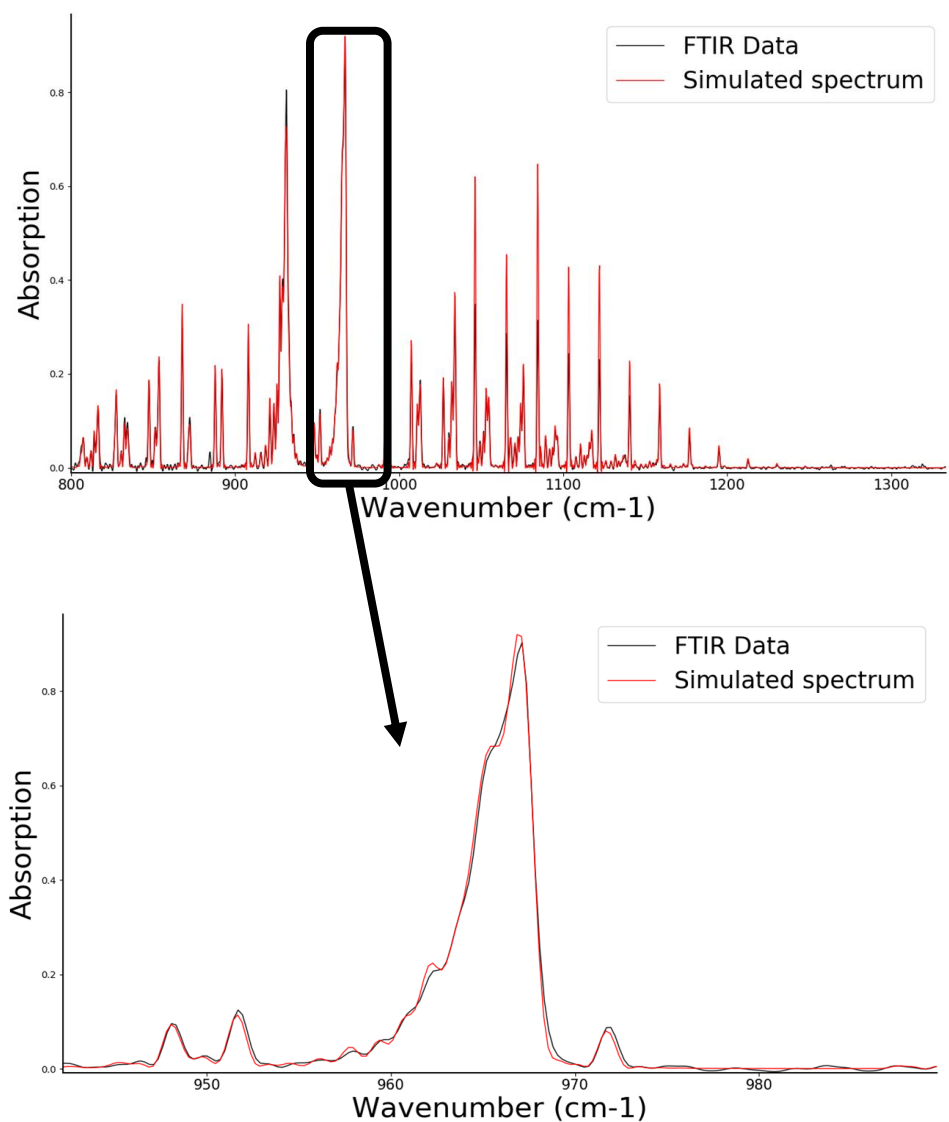

**Figure 3.** Representative spectrum from a SBA-15 reactor with silver.

We determined scaling by integrating a small region (20cm$^{-1}$ in width) in the measured spectrum. To maximize the signal-to-noise ratio, we chose a region centered on the highest peak in the spectral region (967 cm$^{-1}$). We integrated over the same region of the simulated spectrum and used the ratio of these areas to determine the scaling factor. To validate the result, we expanded the window to 200cm$^{-1}$ in width and integrated both spectra again. For all spectra reported here as results, the new ratio corresponded to the original within 5%. (See **Figure S7** for FTIR spectra temperature dependance)

A second validation was performed using known concentrations of ammonia (100 ppm, 500 ppm, and 1000 ppm) supplied by Airgas. **Figure 4** shows the measured concentration, calculated with the scaling factor described above, compared to the calibration concentration stated by the manufacturer.

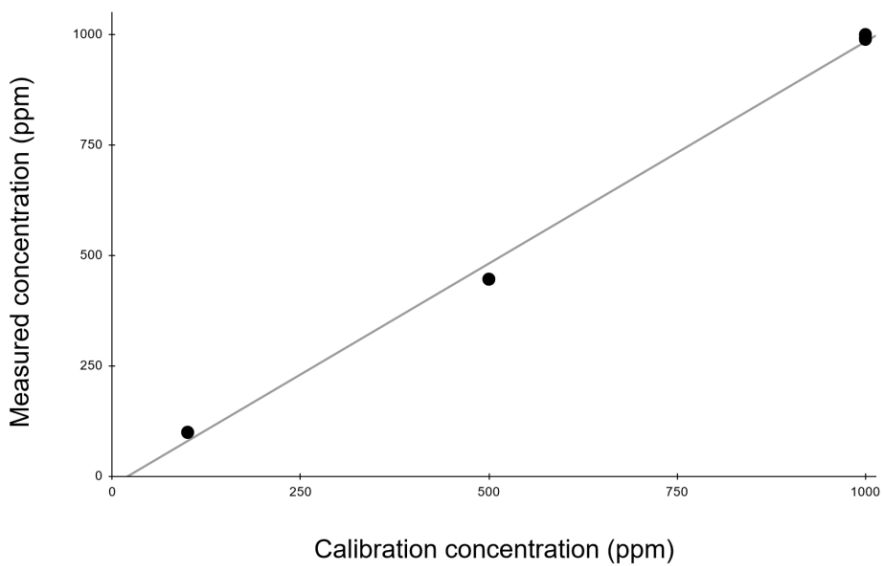

**Figure 4.** Measured concentration vs calibration concentration.

*Optical Emission Spectroscopy*

OES was used to estimate rotational temperature, electron density, vibrational excitation temperature as well as to detect precursor species, NH and H in the discharge. Low resolution survey spectra were taken using Ocean Optics USB4000 (Ocean Insight), 375 nm – 700 nm, and USB2000, 290 – 500 nm range. The spectrometers were calibrated using a black body, xenon light source. High resolution spectra were measured using a Chromex-250 IS (Princeton Instruments), a 25 cm monochromator with 2400 g/mm, 40 μm slit width. The image was recorded by Andor iStar ICCD (Oxford Instruments). 50 on-CCD accumulations were taken for most measurements and the intensifier gain was adjusted in the range of 100 to 200 to achieve the best signal to noise ration while avoiding saturation. A central region of the discharge was imaged onto the entrance slit of the monochromator using one broad band lens as shown in **Figure 1**. The position of the lens was adjusted to maintain focus in each spectral range. Oriel Xe and Ar-Hg calibration lamps were used for wavelength calibration and for fitting the instrument function for the Chromex monochromator.

Specair (SpectraFit, S.A.S.) software package based on the collisional radiative model of air plasma emission[61] was used for temperature estimates. The model contains $N_2$, $N_2^+$, N, NH, and other transitions relevant to this investigation. The $N_2$ spectra contain optical emission corresponding to electronic transitions between excited electronic-vibrational states with rotational sublevels. The rotational sublevels usually well thermalized at atmospheric pressure and are in good equilibrium with the gas translational energy and hence are often used as an estimate of the gas temperature. The emission corresponding to the $N_2$ second positive system, $C^3\Pi_u - B^3\Pi_g$ transition, was used for the estimates below. The 0-0 transition with the band head at 337.13 nm was used for estimations of the rotational excitation temperature.

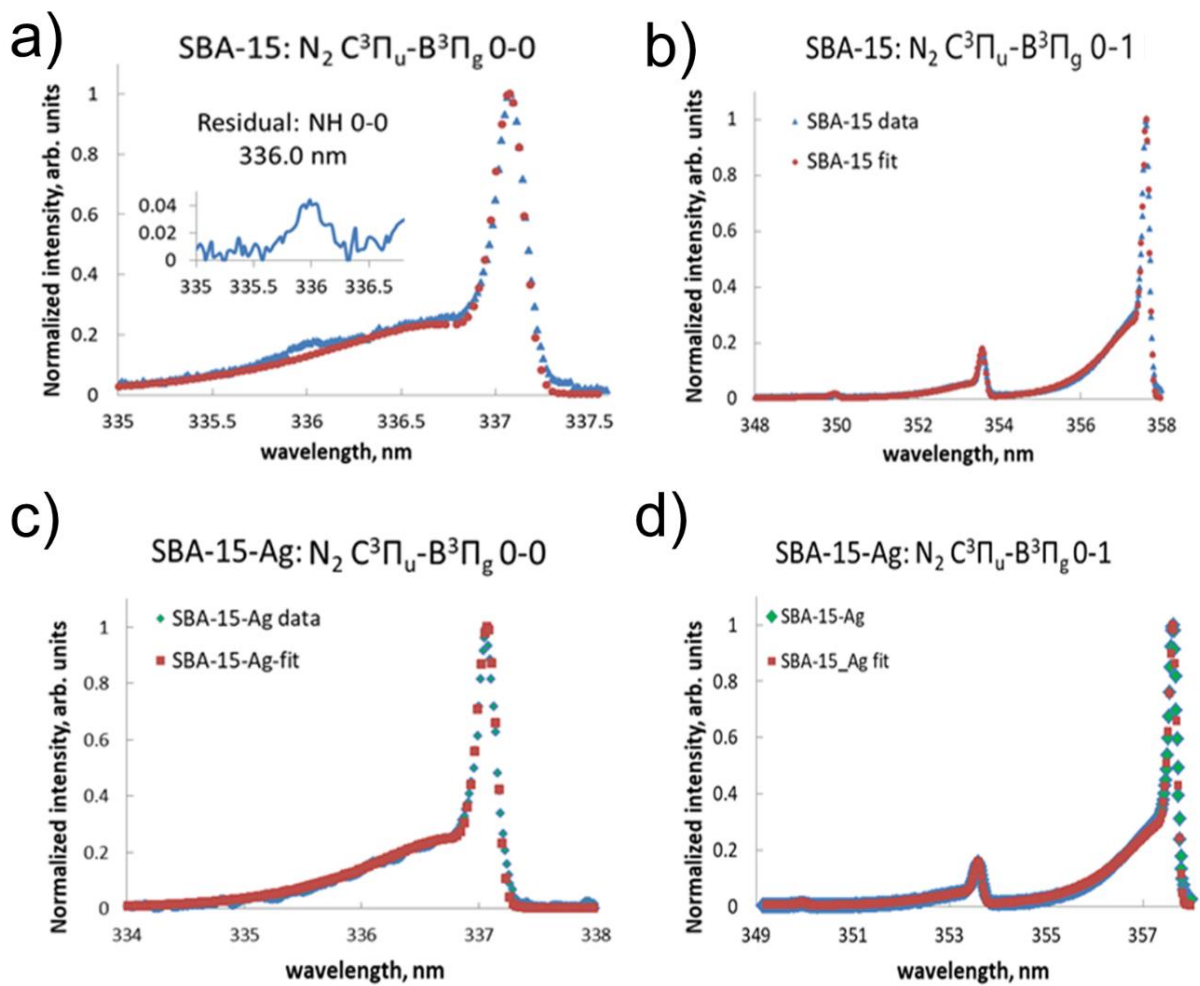

**Figure 5.** N$_2$ emission spectra corresponding to the $C^3\Pi_u - B^3\Pi_g$ 0-0 and 0-1 transitions used for the diagnosis of the vibrational and rotational excitation temperatures of N$_2$ gas. Specair was used to generate and optimize the fitted spectrum, as shown in the figures above. a) SBA-15 N$_2$ (0-0) band, b) SBA-15 N$_2$ (0-1) band, c) SBA-15-Ag N$_2$ (0-0) band, d) SBA-15-Ag N$_2$ (0-1) band,

Specair was used to simulate and fit the experimental spectrum with spectra for a variety of rotational excitation temperatures and the best fit was chosen for the reported temperature (**Figure 5**). This temperature was then used as the initial guess for a fitting process for the +1 (0-1, 1-2, 2-3) vibrational transitions with band heads at 357.69 nm, 353.67 nm, and 350.05 nm respectively. Fitting of this band was performed with vibrational excitation temperature as the varying parameter and the rotational temperature from the fit for the 0-0 transition, 337.13 nm.[62] A simulated spectrum for the same +1 vibrational band was also produced by optimizing both vibrational and rotational temperatures. The results of these fitting processes were close, and any discrepancies were used to evaluate the uncertainties in the temperature estimates.

The average electron density was estimated for each reactor, SBA-15 and SBA-15-Ag using the Stark broadening of $H_\alpha$ emission line profile (**Figure 6**). $H_\beta$ is usually preferred due to its higher sensitivity to Stark broadening[63], but $H_\beta$ emission was not detected in our spectra. At electron densities of the order of $10^{15}$ cm$^{-3}$ or greater, the broadening of the $H_\beta$ line in addition to its lower intensity can make it difficult to discern it above the noise level. According to[64] using $H_\alpha$ may lead to an overestimation of the electron density of up to 35% compared to $H_\beta$ and $H_\gamma$, but the accuracy is better at densities $>10^{15}$ cm$^{-3}$ and this line has been used to calculate electron density by several groups.[65, 66, 67] In atmospheric pressure plasmas the line profile can be approximated by a Voight profile, $f(\lambda)$, which is a convolution of the Gaussian and Lorentzian profiles:

$$f(\lambda) = \frac{2ln2 \Delta \lambda_L}{\pi^{3/2} \Delta \lambda^2{}_G} \int_{-\infty}^{+\infty} \frac{e^{-t^2} dt}{\left[2\sqrt{ln2}\frac{(\lambda-\lambda_0)}{\Delta \lambda_G} - t\right] + \left[\sqrt{ln2}\frac{\Delta \lambda_L}{\Delta \lambda_G}\right]^2},$$

Where $\Delta \lambda_L$, $\Delta \lambda_G$, are the FWHM of the Gaussian and Lorentzian profiles. Several fitting procedures were conducted varying both Gaussian and Lorentzian profiles, but due to noise and finite sampling rate, these procedures did not produce meaningful results or failed converge. Therefore, one parameter fitting was performed where the Gaussian broadening was fixed, and only the Lorentzian broadening was used as a fitting parameter. Similar difficulties were described in[63]. Gaussian broadening was determined as

$$\Delta \lambda_G = \sqrt{\Delta \lambda_D{}^2 + \Delta \lambda_i{}^2},$$

where $\Delta\lambda_D$, is the Doppler broadening and $\Delta\lambda_I$, is the instrumental function. The instrument function was determined using a Hg-Ar calibration lamp, to be $\Delta\lambda_I = 0.15\ nm$. The Doppler broadening was estimated as[68, 69]

$$\Delta\lambda_D = 7.16 \cdot 10^{-7} \cdot \lambda \cdot \sqrt{\frac{T_g}{M}},$$

where $\lambda = 656.279\ nm$, M = 1 g/mol is the atomic mass of H, and $T_g$ is the temperature of the H atoms, which is considered equal to the gas temperature which is in turn estimated as the rotational excitation temperature of $N_2$ molecules. The highest $T_{rot}$ estimate in our experiment is 500 K (see above for the methods and the results section for the specific estimates), so $\Delta\lambda_D \leq 0.01\ nm$ introduces only a small correction to the instrument function, $\Delta\lambda_I = 0.1503\ nm$.

Fixing the Gaussian broadening at this value and fitting the Voigt profile to the line with Lorentzian broadening as a fitting parameter gives the Lorentzian broadening of 0.28 nm for SBA-15 and 0.178 nm for SBA-15-Ag. This is comprised of the pressure and Stark broadening, $\Delta\lambda_L = \Delta\lambda_P + \Delta\lambda_S$. In our case of 1:1 molar ratio of $N_2$ to $H_2$ and atmospheric pressure, we consider both types of pressure broadening, resonance and Van der Waals broadening[69]). Resonance broadening of H[70]

$$\Delta\lambda_R = 8.6 \times 10^{-28} \cdot \left(\frac{g_i}{g_k}\right)^{1/2} \lambda^2 \lambda_r f_r N_g,$$

where $\lambda$ is the wavelength of the observed line, $f_r$ and $\lambda_r$ are the oscillator strength and wavelength of the resonance line; $g_i$ and $g_k$ are the statistical weights of its upper and lower levels, $N_g$ is the ground state number density. For $L_\beta$ resonance line, $\lambda_r = 102.572\ nm$, and $f_r = 0.0791$, $g_i = 2$ and $g_k = 18$. The number density for atomic hydrogen is difficult to assess but estimating the Boltzmann factor for the assumed electron temperature of ≤1 eV and the dissociation energy of 4.52 eV gives ~0.01, so for a partial pressure of ½ atm, $N_g \sim 10^{17} cm^{-3}$. The resonance broadening $\Delta\lambda_R \sim 10^{-4}$ nm and can be neglected.

Next, we consider the Van der Waals broadening of the $H_\alpha$ line, which is given by

$$\Delta\lambda_{vdW} = 2.06 \times 10^{-13} \cdot \lambda^2 (\alpha \underline{R}^2)^{\frac{2}{5}} \left(\frac{T_g}{\mu}\right)^{3/10} N_n,$$

where α is the polarizability of the perturber ($1.8 \times 10^{-30} m^3$ for nitrogen gas[70]), μ is the reduced mass (kg), $N_n$ is the neutral number density ($m^{-3}$) and $\underline{R}^2$ is the difference between the values of the square radius of the emitting atom in the upper k and lower level i. The Van der Waals broadening (nm) of $H_\alpha$ due to the nitrogen gas from[67]:

$$\Delta\lambda_{vdW} \approx 0.1 \cdot \frac{p_N}{\left(\frac{T_g}{300}\right)^{0.7}},$$

Where $p_N$ is partial pressure of nitrogen gas in bar, and $T_g$ is the gas temperature estimated as the rotational excitation temperature of the $N_2$ vibrational bands as described above. For example, for $T_g = 500\ K$, $\Delta\lambda_{vdW} \approx 0.035$ nm. Subtracted from the total Lorentzian broadening obtained from fitting the line profile, it gives the Stark broadening that results from the Coulomb interaction of the emitting atoms (here H) with the charged particles in the plasma. Electrons are responsible for most of this broadening due to their higher velocities. Full width at half maximum for Stark broadening component is given by

$$\Delta\lambda_S = 2.5 \times 10^{-10} \cdot \alpha(n_e, T_e) \cdot n_e^{2/3},$$

Where $n_e$ and $T_e$ are the electron density and temperature in cm$^{-3}$ and K, respectively, and $\alpha(n_e, T_e)$ is the reduced wavelength separation tabulated in detail for a wide range of temperatures in[66] demonstrating the validity of the approximation given in[65]

$$\Delta\lambda_S = 1.0989 \cdot \left(\frac{n_e}{10^{17}}\right)^{0.67965}$$

This approximation gives the $\Delta\lambda_S$ in the range 0.15 – 0.25 nm for electron density in the range of $10^{15}$ – $10^{16}$ cm$^{-3}$ with about 5% temperature dependent variation for in the 10,000 K range and 10% variation for temperatures in the 20,000 K to 40,000 K.[65, 66]

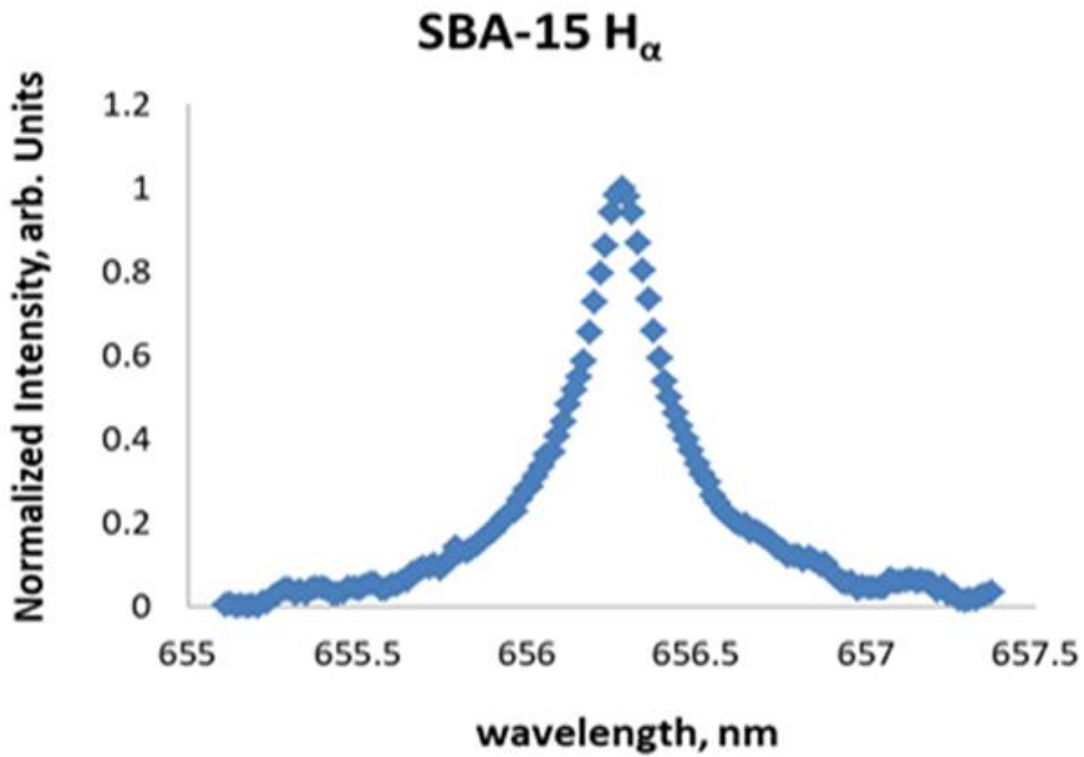

**Figure 6.** Normalized H$_\alpha$ emission line for SBA-15 at the following operational parameters 11.2±0.2 kV p-p, 10.7 ± 0.9W, 44.4 ± 0.4 kHz

*Monitoring ammonia synthesis using Fourier Transform Infrared Spectroscopy*

Experiments were first carried out to test three selected metal catalysts acting as electrodes. Plug reactors with silver, palladium, and nickel central electrodes were tested. For each electrode, we tested two feed gas ratios $N_2$:$H_2$ of 1:1 and 1:3 both at 25 sccm total flow rate. Prior to each trial, the reactor was flushed for 20 minutes with the same gases as were used during operation. After flushing, the power supply was turned on and adjusted to the reactor's resonant frequency to maximize the voltage applied to the reactor and a minimum current from the power source. Silver was selected as the metal catalyst for the packed bed reactors due to its better performance for ammonia synthesis. Non-stoichiometric gas ratio of 1:1 $N_2$:$H_2$ was used in experiments with the packed bed reactors.

For the second stage of the experiment, we investigated the effect of silver on ammonia synthesis in a packed-bed reactor. Mesoporous silica SBA-15, with 7 nm diameter pores was synthesized as the packing material for the reactor as well as SBA-15-Ag, with 10% wt. of Ag.

To remove adsorbed water and clean the reactors from remaining residue from the SBA-15 synthesis process and/or from previous experiments, the reactors were pre-treated in $H_2$ plasma.[71, 72] $H_2$ plasma is chosen due to its low atomic mass to prevent sputtering of SBA-15 by inert gases and nitrification in $N_2$ gas during plasma treatment although it leaves the possibility of hydrogenation particularly of the metal sites in SBA-15-Ag. Each reactor was flushed with $H_2$ gas for 20 minutes at 25 sccm/min (which corresponds to the total gas flow rate in all experiments), and then plasma was ignited while maintaining the flow rate. Power was then applied to the reactor at a 5 kV maximum voltage for one hour. FTIR spectra of the output gas were taken during the $H_2$ plasma treatment, showing an immediate increase in water vapor concentration downstream, which then decayed over time, indicating that water was removed from the packing material.

After one hour of plasma processing the reactor, the power was turned off while the gas continued to flow allowing the reactor to cool down. After cooling for several minutes, the gas flow rate of $H_2$ was gradually reduced while increasing the flowrate of nitrogen gas $N_2$ while keeping the total flow rate constant. Ammonia synthesis experiments were performed in 1:1 $N_2$:$H_2$ plasma in packed bed reactors, filled with SBA-15 or SBa-15-Ag. The synthesis methods for these materials are described in **Section 2**.

In each synthesis trial, FTIR spectra were acquired at intervals of roughly four minutes. The resulting ammonia concentrations over time are shown in **Figure 7**. For all trials, there was a delay of about five minutes between starting the reactor and observing any ammonia in the measurement cell. We attribute this to the gas travel time over the approximately 0.5m distance from the reactor output to the cell. After this time, a period of faster increase followed, and the concentration then leveled out from approximately 40 to 60 minutes of operation. We report ammonia production for each reactor using the average concentration within this "steady state" period.

The operating conditions for packed bed reactors were chosen such that a reactor exhibited similar performance across repeated trials, i.e. an approximately one-hour trial did not within experimental error of each trial, affect ammonia production in subsequent trials. Above an applied voltage of about $V_{p-p}$ = 16 kV, performance for packed-bed reactors was found to be non-repeatable. The packing material changed in color during the trial at this voltage, and subsequent trials yielded significantly different ammonia production rates. These effects may warrant an additional investigation that would involve monitoring the chemical and morphological changes in the packing material during operation and track its effects on ammonia production.

Plug reactors produce much lower concentrations of ammonia at the same applied power and so were operated at higher power for ease of ammonia detection. Plug reactors produced repeatable concentrations at all voltages in our experiments. Although different applied powers affect the quantitative results, they do not affect the qualitative conclusions of this work. It is difficult to compare operational conditions of plasma sources of different nature such as plug reactors and packed bed reactors. Hence, operational conditions were selected independently for each type of reactor and kept consistent across experiments with the same type of reactor.

For the packed bed reactors, the applied voltage was set to $V_{p-p}$ = 11 kV and the duty cycle was 40% except for the higher power tests that were conducted at at $V_{p-p}$ = 14kV.

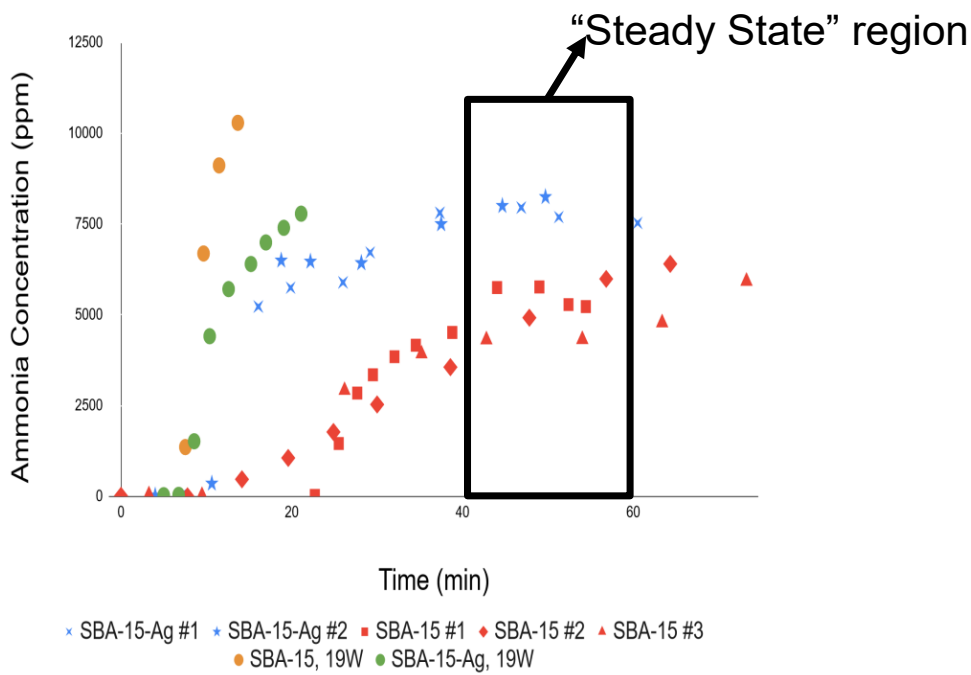

**Figure 7.** Ammonia concentration as a function of time for different catalysts.

*Electrical Characterization*

Electrical characteristics were monitored during all trials to determine the operating power of the reactor. A typical oscilloscope trace, corresponding to a SBA-15-Ag packed-bed reactor, is shown in **Figure 8**. Although the power supply frequency was adjusted to resonance in every trial, it varied little between reactors. For the plug reactors, the average frequency was 41.7 +- 0.2 kHz and for the packed-bed reactors it was 44.4 +- 0.4 kHz.

The plug reactors have a cylindrical geometry with the quartz tube acting as the dielectric barrier. The electrical parameters were not sensitive to the electrode material for Ni, Pd, and Ag electrodes in this study. Typical current, voltage, and charge traces are shown in **Figure 8**. Lissajous figures indicate the changes in the electrical properties of the reactors during operation. The shape of the Lissajous figures, traditionally a parallelogram, is interpreted to indicate the capacitance of the system during the discharge off and on times, $\frac{dQ}{dV} = C$, where Q is charge, V is voltage between the plug high voltage and the mesh ground electrode, and C is the capacitance of the entire system at various applied voltages during operation. For the plug reactors,

$$\frac{1}{C} = \frac{1}{C_{tube}} + \frac{1}{C_{gas}},$$

where $C_{tube}$ is the geometrical capacitance of the quartz tube and $C_{gas}$ is the capacitance of the air gap. During the discharge on time, the gas gap is bridged, and the capacitance increases to $C_{tube}$, the capacitance of the quartz tube. Here the estimations of the capacitance from the straight portions of the Lissajous figures, $C_{on} = 40.4 \pm 1.2\ pF \approx C_{tube}$ agrees with the geometrical estimate of the capacitance of a quartz tube, and likewise $C_{off} = 8.8 \pm 1.2\ pF \approx C$, the capacitance of the entire reactor. The deviations from the parallelogram shape have been attributed to the discharge spreading to cover the electrode, hence $C_{on}$ represents the maximum value of the system capacitance during discharge.

The packed-bed reactors were operated at $V_{p-p}$ = 11kV, for the majority of ammonia synthesis experiments described here, and the power consumption was $P_{ave}$= 10.7 +- 0.9 W. We also ran these reactors at a higher power with $V_{p-p}$ = 14kV, $P_{ave}$= 17.3 +- 1.0 W. The Lissajous figures demonstrate that the addition of 10% wt. Ag to the mesoporous packing material doesn't have any significant effect on the electrical properties of the reactors. The shape of the Lissajous figures is different from that for plug reactors, with a greater distortion and a lower maximum capacitance of the reactor during the plasma on time. These differences can be explained by the presence and the structure of the packing material. The overall capacitance of the packed bed reactor should be lower since the reactor capacitance can be approximated as,

$$\frac{1}{C} \approx \frac{1}{C_{tube}} + \frac{1}{C_{gas}} + \frac{1}{C_{pack}},^{73}$$

where $C_{pack}$ is the capacitance of the packing material. This expression is only approximate and to date there is no adequate model of a packed bed reactor. The approximate expression does provide some insight into the operation of this reactor. The observed $C_{off} = 7.7 \pm 1.2 \, pF \approx C$, is only slightly lower than for a plug reactor which can be attributed to a very low dielectric constant of mesoporous silica packing material $\varepsilon_r < 2.^{74}$ The difference between the reactor capacitances is much greater during the discharge on time with $C_{on} = 25.3 \pm 3.4 \, pF$ indicating that the gap between the center electrode and the quartz tube is not bridged entirely by the discharge. As described previously [46, 47] there is a large proportion of partial discharges in a packed bed reactor, discharges between the individual beads due to the polarization and charge accumulation on the surface of the packing material. These microdischarges are responsible for the irregularities observed in the Lissajous figures of the packed reactors. The smoother curved part of the Lissajous figures may still correspond to the spreading of the discharge although now the surface covered is not the electrode but the individual beads in the reactor.[58, 73]

In summary, electrical characteristics illuminate significant differences between the plug and packed reactors and little or no difference between packed reactors with and without metal catalyst.

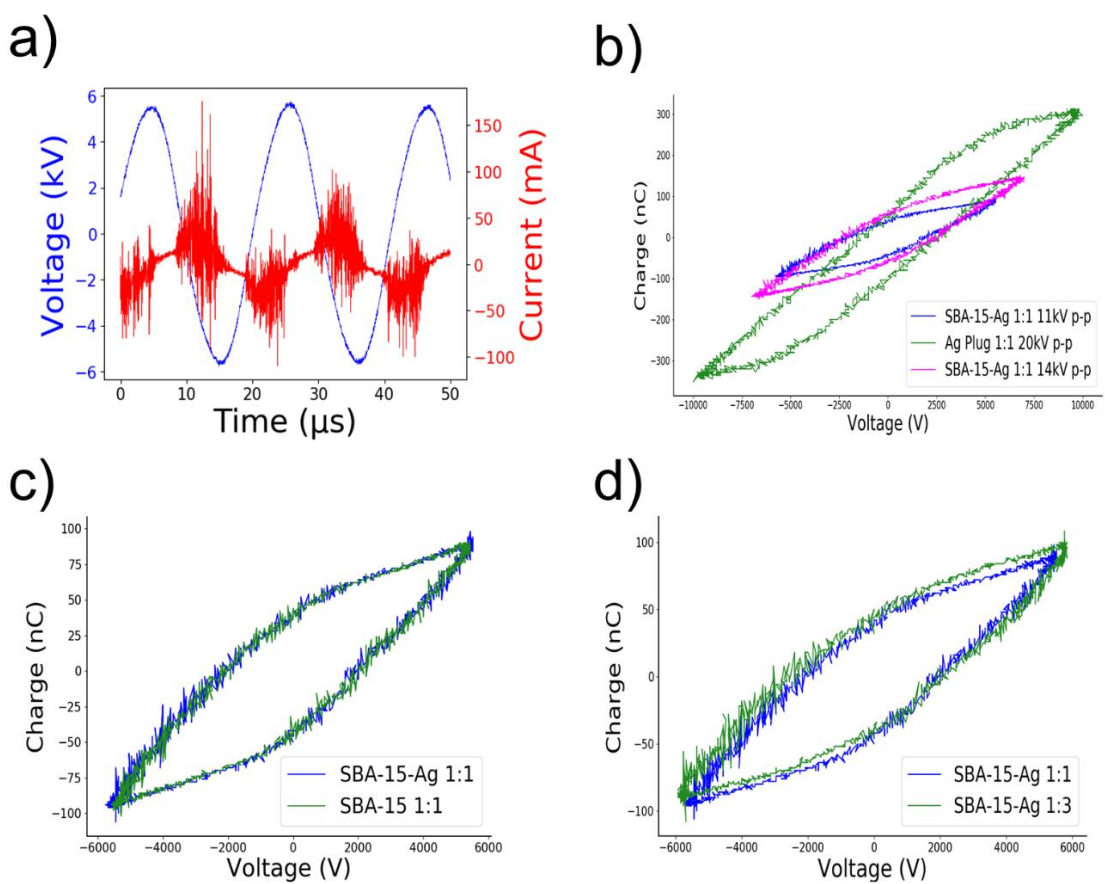

**Figure 8.** Electrical characteristics of the plug and packed reactors for ammonia synthesis, where a) Voltage, Current, and Charge traces for a typical reactor, b) Lissajous plot for the plug reactor with Ag electrode and packed bed reactors SBA-15-Ag, c) and d) demonstrate the consistency of the discharges in packed reactors with and without a metal catalyst (c) and for two different $N_2:H_2$ gas ratios d).

*Ammonia synthesis with transition metal catalysts*

Our experimental strategy was to select a metal catalyst for use with plasma assisted catalysis and test this metal catalyst in a packed bed reactor with a mesoporous catalyst to access both, surface dominated, and metal assisted regimes.

Ammonia synthesis experiments were performed with plug reactors using silver, palladium, and nickel using $N_2:H_2$ gas ratios of 1:1 and 1:3 and ammonia concentrations were determined using the FTIR spectroscopy methods discussed above.

Results for all six conditions are presented in **Figure 9**. Ni and Pd produced low concentrations of ammonia, <200 ppm and ~200 ppm respectively. The results for different gas ratios of $N_2:H_2$ are the same within experimental error. The reactor with the Ag electrode produced a significantly higher concentration of ammonia than that of Ni and Pd. The gas ratio has a significant effect on the ammonia production with Ag as the catalytic material. The highest synthesis rate was observed with the silver electrode at 1:1 $N_2:H_2$ ratio. Based on these results, we chose the gas ratio of 1:1 $N_2:H_2$ for the packed bed reactors, and Ag as the metal catalyst to be added to the mesoporous silica, SBA-15, powdered packing.

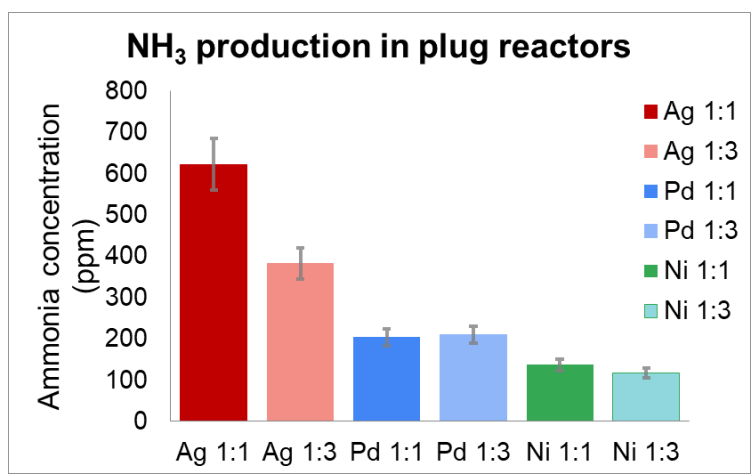

**Figure 9.** Ammonia production in plug reactors.

*Ammonia synthesis in packed bed reactors*

Overall in our experiments, the packed bed reactors produce at least an order of magnitude higher concentration of ammonia at lower operating voltage and power in agreement with the findings by Carreon et al. and others.[44, 56, 75] At higher power of 19 W, applied voltage, 14 kV p-p, SBA-15 without Ag produced 10,300 ppm of ammonia or about 24% higher concentrations of ammonia than SBA-15-Ag of 7,700 ppm (**Figure 10**). At lower power of 11 W and 11 kV p-p, the production trend is reversed with SBA-15-Ag producing 7900 ppm, a 33% higher concentration than SBA-15 of 5300 ppm.

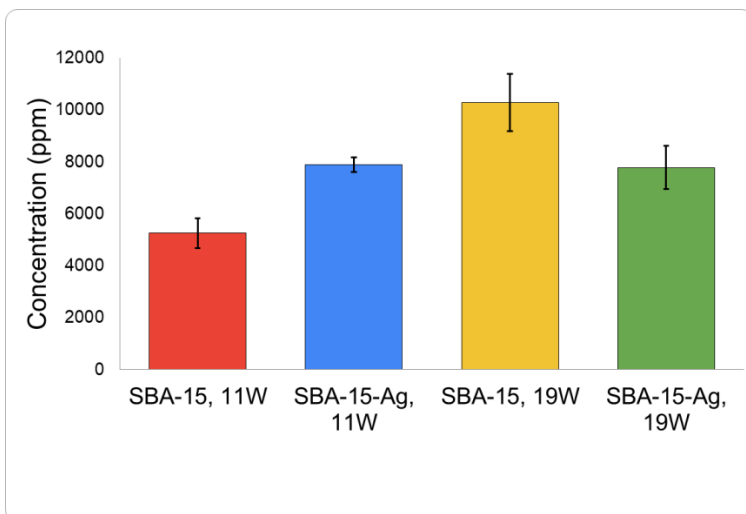

**Figure 10.** Ammonia production in packed bed reactors.

*Plasma characterization using optical emission spectroscopy.*

The low-resolution survey spectrum is dominated by the $N_2$ emission bands. High resolution spectra in the 335 – 337 nm range, features the 0-0 band of the $N_2$ second negative rotation-vibrational system with the head at 337.15 nm for both SBA-15 and SBA-15-Ag (**Figure 5**) The NH 0-0 vibrational band at 336.0 nm was detected in the spectra of SBA-15 only and not SBA-15-Ag. Although emission band intensities are affected by many factors including plasma density and temperature, the gas and surface temperatures in the reactor, etc., the NH feature is not present when the Ag is added to the silica. In addition, the $H_\alpha$ emission is much more prominent in the spectrum of SBA-15.

$N_2$ emission bands were used to determine rotational and vibrational excitation temperatures. These differ only slightly for the two compositions, SBA-15 and SBA-15-Ag. Rotational excitation temperature is $T_{rot}= 500 \pm 20$ K for SBA-15 and 50 K higher for SBA-15-Ag which may be attributed to the localized heating around the metal sites.[76, 77] On the other hand, vibrational temperature is slightly higher for silica without silver, 2100 K, than with silver, 1900 K. Although these two parameters are close, OES reveals other differences between the plasma reactors with these two materials.

The characteristic NH emission band at 336 nm overlaps with the $N_2$ 337.15 nm band and hence it is difficult to detect, and it can affect the spectral fitting and rotational temperature estimates from the 337.15 nm band. The spectra from the reactors without Ag, have a distinct feature at 336 nm and subtracting the simulated spectrum has a band-like residual at 336 nm (**Figure 5**). This indicates the presence of NH in gaseous phase.

**Figure 11** shows the spectra of reactors with SBA-15 and SBA-15-Ag in the 652 nm – 660 nm range. This spectral region is dominated by the $N_2$ emission bands, but the hydrogen $H_\alpha$ line at 656.28 nm is clearly visible. The intensity of the $H_\alpha$ relative to the $N_2$ emission bands is much greater for the reactor without Ag as seen from the superimposed scaled spectra (**Figure 11**). The band at 391.4 nm is identified as the 0-0 band of the first negative system at characteristic to $N_2^+$ ion emission (**Figure 12**). The chosen spectral window also includes several characteristic bands of the neutral $N_2$ emission that have a high probability of excitation from ground state and therefore have intensities proportional to the concentration of neutral nitrogen. The ratio of the intensity of the 391.4 nm band to the 399.4 nm 1-4 transition of the second positive system of $N_2$ indicates the degree of ionization and hence the plasma density. The spectra in **Figure 12** have been normalized to the highest intensity in this region, the 391.4 nm $N_2^+$ line and show a higher relative intensity of the ionized $N_2^+$ versus the neutral $N_2$ emission. This spectral evidence indicates a higher plasma density in the reactors without the added Ag.

To assess the plasma density in the reactors quantitatively we used the Stark broadening of the $H_\alpha$ line as described in detail in the methods section above. The plasma density of the reactors without a metal catalyst, $N_e \approx 9 \times 10^{15}$ cm$^{-3}$, which is 45% higher than plasma density in the reactors with added Ag where $N_e \approx 5 \times 10^{15}$ cm$^{-3}$. The estimated values of all plasma parameters evaluated here are summarized in **Table 2**.

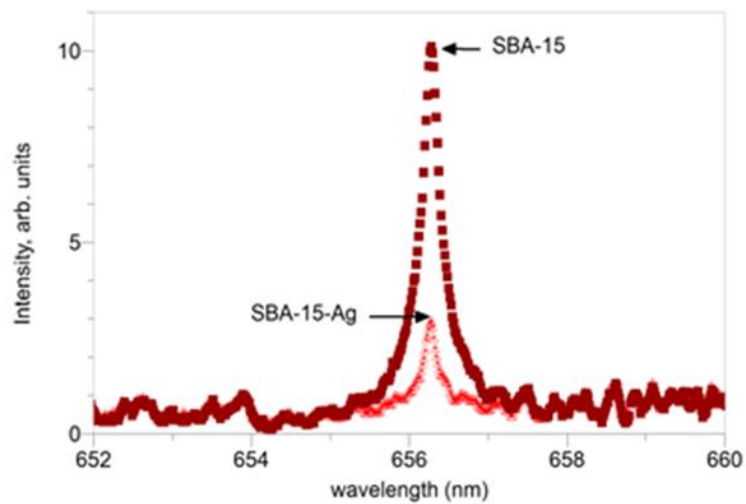

**Figure 11.** Intensity of the $H_\alpha$ relative to the $N_2$ emission bands.

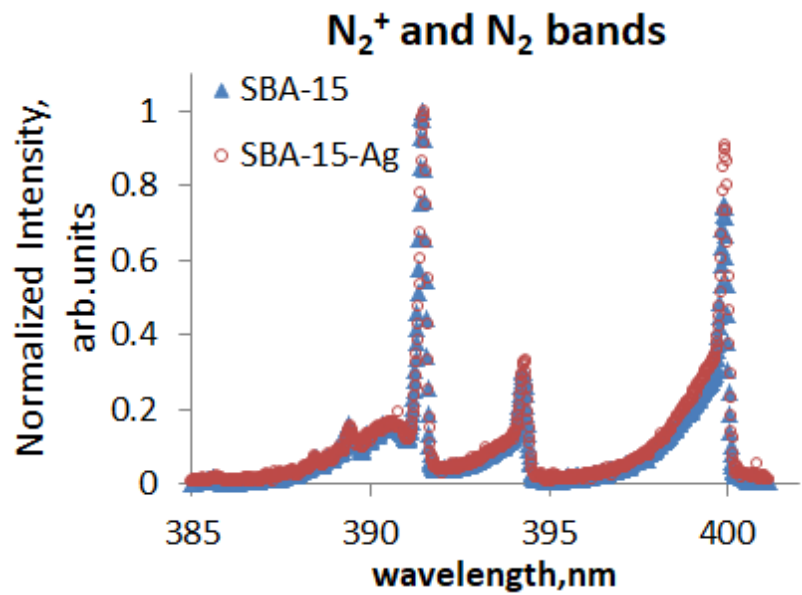

**Figure 12.** Nitrogen spectra.

**Table 2** OES estimations of plasma parameters, rotational and vibrational excitation temperatures and electron density Ne for mesoporous silica SBA-15 and mesoporous silica with 10% wt. Ag, SBA-15-Ag, packed bed reactors. Operational conditions were 11 W and 11 kV p-p.

**Table 2.** SBA-15-Ag Vs. SBA-15, rotational and vibrational excitation temperatures, and electron density

|  | **SBA-15-Ag** | **SBA-15** |
|---|---|---|
| $T_{rot}$, K | 550±20 | 500±20 |
| $T_{vib}$, K | 1900±100 | 2100±100 |
| $N_e$, cm$^{-3}$ | $5 \times 10^{15}$ | $9 \times 10^{15}$ |

**Conclusions**

The presented data indicate the possibility that certain discharge conditions can lead to separate regimes, potentially as a result of different interactions of plasma generated species with metals and porous materials, respectively. It is clear from the presented results the possible existence of different reaction mechanisms when employing different packing materials due to the change in the interactions between plasma generated species and the catalytic surfaces. However, a transition boundary between different existing regimes remains to be delineated. We anticipate that the understanding and control of the boundary evolution for different catalyst designs and plasma conditions will contribute to our mechanistic understanding of this electron activated catalytic process. This will result in the design of economical plasma intermittent small-scale decentralized systems to produce ammonia. Below we summarize key points of our findings in this work:

1. The catalysts required for plasma catalysis are different than thermal catalysis because the rate limiting nitrogen dissociation process is largely taken care of by plasma.
2. We performed the selection of a metal catalyst from a group of transition metals by using them in plug reactors first. These are dielectric barrier reactors which are very good at nitrogen excitation and possibly, dissociation, especially at higher powers. We then used the best performing metal dispersed on mesoporous silica SBA-15 in a packed bed reactor. The use of plug reactors for the selection of a catalyst has the advantages of plasma environment and limited surface effects but there are also drawbacks. Although the plug reactors have extremely small surface area compared to the packed bed reactors, the presence of the metal electrode has shown an impact in the ammonia production while the dependence with the type of metal is modest.[78,56] With the possibility that a major percentage of the reactions in plasma occur on the walls.[56] In addition, the plasma conditions are different in the packed bed reactors[44] and the selected catalysts may perform better in one reactor but not the other.

3. We have demonstrated that the performance of the silica scaffold and the metal catalyst depend on the plasma conditions. At higher power, mesoporous silica without Ag outperforms SBA-15-Ag, producing the largest concentration of ammonia observed in our experiments. At lower power, the addition of Ag results in 33% higher concentration of ammonia. To understand this result we use OES to compare the reactors with and without metal catalyst.

4. OES data show that SBA-15 has higher electron density. Since we couldn't employ atomic lines to determine the electron temperature, we made an assumption that the electron temperature is in the 0.7 – 2 eV range, a reasonable assumption for a packed bed reactor at atmospheric pressure.[72, 79] The parameter tables in Grigosos et al[65] for the $H_\alpha$ profiles show that the temperature dependence affects plasma density by less than 10% for this temperature range. Higher electron density may result in higher dissociation of $N_2$ and hence access plasma assisted surface dominated synthesis regime.

5. The distortions in the Lissajous figures for the packed reactors are an indication of point-to-point partial discharges which may contribute higher electron density. Stronger $N_2^+$ emission provides additional evidence for higher electron density in the SBA-15 reactors.

6. The addition of Ag does not affect the electrical properties of the discharge and result in heating the gas by 50 K. The $T_{vib}$ is higher for the SBA-15 reactors. It is not clear if the slight increase in vibrational temperature can be significant for the synthesis reactions.

7. The addition of Ag lowers the concentration of atomic hydrogen. Where Ag acts as a hydrogen sink facilitating surface reactions with nitrogen. NH in the spectrum may indicate a higher concentration of H in the SBA-15 reactor.

ASSOCIATED CONTENT

**Supporting Information**.

Details on SEM images for SBA-15 and SBA-15-Ag, Details on morphology before and after plasma exposure, Details on TEM images for SBA-15-Ag, Details on pore size distribution for SBA-15 and SBA-15-Ag, Details on particle size distribution for SBA-15-Ag, Details on FTIR spectra (Temperature dependence).

Corresponding Author

*Corresponding author: *maria_carreon@uml.edu*

**Acknowledgements**

The work by SG and HF is supported by the Princeton Collaborative Research Facility (PCRF), which is supported by the U.S. Department of Energy (DOE) under Contract No. DE-AC02-09CH11466. The work by MLC and FG is supported by DOE, Office of Science Fusion Energy Sciences Award No. DE-SC0021309. All plasma diagnostic resources used in this work were provided by the PCRF. MLC wants to thank Dr. Jacek Jasinski for his help with the collected TEM images and  insightful discussion about the same.

**References**

1. Reiter, A. J.; Kong, S.-C., Combustion and emissions characteristics of compression-ignition engine using dual ammonia-diesel fuel. *Fuel* **2011,** *90* (1), 87-97.
2. Carreon, M. L., Plasma Catalytic Ammonia Synthesis: State of the Art and Future Directions. *Journal of Physics D: Applied Physics* **2019,** *52* (48), 483001.
3. Degnan, T., New catalytic developments may promote development of smaller scale ammonia plants. *Focus on Catalysts* **2018,** *1*.
4. Klerke, A.; Christensen, C. H.; Nørskov, J. K.; Vegge, T., Ammonia for hydrogen storage: challenges and opportunities. *Journal of Materials Chemistry* **2008,** *18* (20), 2304-2310.
5. Rees, N. V.; Compton, R. G., Carbon-free energy: a review of ammonia-and hydrazine-based electrochemical fuel cells. *Energy & Environmental Science* **2011,** *4* (4), 1255-1260.
6. Wang, W.; Herreros, J. M.; Tsolakis, A.; York, A. P. E., Ammonia as hydrogen carrier for transportation; investigation of the ammonia exhaust gas fuel reforming. *international journal of hydrogen energy* **2013,** *38* (23), 9907-9917.
7. Zamfirescu, C.; Dincer, I., Ammonia as a green fuel and hydrogen source for vehicular applications. *Fuel processing technology* **2009,** *90* (5), 729-737.
8. Lan, R.; Irvine, J. T. S.; Tao, S., Ammonia and related chemicals as potential indirect hydrogen storage materials. *International Journal of Hydrogen Energy* **2012,** *37* (2), 1482-1494.
9. Cortright, R. D.; Davda, R.; Dumesic, J. A., Hydrogen from catalytic reforming of biomass-derived hydrocarbons in liquid water. *Nature* **2002,** *418* (6901), 964-967.
10. Christensen, C. H.; Johannessen, T.; Sørensen, R. Z.; Nørskov, J. K., Towards an ammonia-mediated hydrogen economy? *Catalysis Today* **2006,** *111* (1-2), 140-144.
11. Erisman, J. W.; Sutton, M. A.; Galloway, J.; Klimont, Z.; Winiwarter, W., How a century of ammonia synthesis changed the world. *Nature Geoscience* **2008,** *1* (10), 636-639.
12. Schlögl, R., Ammonia synthesis. In *Handbook of heterogeneous catalysis*, Wiley-VCH: 2008; pp 2501-2575.
13. Patil, B. S., Plasma (catalyst)-assisted nitrogen fixation: reactor development for nitric oxide and ammonia production. Technische Universiteit Eindhoven: **2017**; pp 1-223.
14. Tanabe, Y.; Nishibayashi, Y., Developing more Sustainable Processes for Ammonia Synthesis. *Coord. Chem. Rev.* **2013,** *257* (17), 2551-2564.
15. http://energy.globaldata.com/media-center/press-releases/oil-and-gas/, global-ammonia-capacity-to-reach-almost-250-million-tons-per-year-by-2018-says-globaldata. 2018.
16. Hong, J.; Prawer, S.; Murphy, A. B., Plasma catalysis as an alternative route for ammonia production: status, mechanisms, and prospects for progress. *ACS Sustain. Chem. Eng.* **2018,** *6* (1), 15-31.
17. Hess, C.; Lercher, J.; Naraschewski, F.; Kondratenko, E.; Baerns, M.; Trunschke, A.; Grasselli., R., *Nanostructured Catalysts: Selective Oxidations*. Royal Society of Chemistry: 2011.
18. Shah, J.; Gorky, F.; Psarras, P.; Seong, B.; Gómez-Gualdrón, D. A.; Carreon, M. L., Enhancement of the Yield of Ammonia by Hydrogen-Sink Effect during Plasma Catalysis. *ChemCatChem* **2020,** *12* (4), 1200-1211.
19. Norby, T.; Widerøe, M.; Glöckner, R.; Larring, Y., Hydrogen in oxides. *Dalton transactions* **2004,** (19), 3012-3018.

20. Efimchenko, V. S.; Fedotov, V. K.; Kuzovnikov, M. A.; Zhuravlev, A. S.; Bulychev, B. M., Hydrogen solubility in amorphous silica at pressures up to 75 kbar. *The Journal of Physical Chemistry B* **2013,** *117* (1), 422-425.
21. Hashizume, K.; Ogata, K.; Nishikawa, M.; Tanabe, T.; Abe, S.; Akamaru, S.; Hatano, Y., Study on kinetics of hydrogen dissolution and hydrogen solubility in oxides using imaging plate technique. *Journal of Nuclear Materials* **2013,** *442* (1-3), S880-S884.
22. Meletov, K. P.; Efimchenko, V. S., Raman study of hydrogen-saturated silica glass. *International Journal of Hydrogen Energy* **2021**.
23. Peng, P.; Chen, P.; Addy, M.; Cheng, Y.; Anderson, E.; Zhou, N.; Schiappacasse, C.; Zhang, Y.; Chen, D.; Hatzenbeller, R.; Liu, Y.; Ruan, R., Atmospheric Plasma-Assisted Ammonia Synthesis Enhanced via Synergistic Catalytic Absorption. *ACS Sustainable Chemistry & Engineering* **2019,** *7* (1), 100-104.
24. Aika, K.-i., Role of alkali promoter in ammonia synthesis over ruthenium catalysts—Effect on reaction mechanism. *Catalysis Today* **2017,** *286*, 14-20.
25. Sugiyama, K.; Kiyoshi, A.; Masaaki, O.; Hiroshi, M.; Tsuneo, M.; ., N. O., Ammonia synthesis by means of plasma over MgO catalyst. *Plasma Chem . Plasma Process .* **1986,** *6* (2), 179-193.
26. Ozaki, A., Development of alkali-promoted ruthenium as a novel catalyst for ammonia synthesis. *Accounts of Chemical Research* **1981,** *14* (1), 16-21.
27. Reddy, R. R.; Ahammed, Y. N.; Gopal, K. R.; Raghuram, D. V., Optical electronegativity and refractive index of materials. *Optical materials* **1998,** *10* (2), 95-100.
28. Holzer, F.; Roland, U.; Kopinke, F.-D., Combination of non-thermal plasma and heterogeneous catalysis for oxidation of volatile organic compounds: Part 1. Accessibility of the intra-particle volume. *Applied Catalysis B: Environmental* **2002,** *38* (3), 163-181.
29. Roland, U.; Holzer, F.; Kopinke, F.-D., Combination of non-thermal plasma and heterogeneous catalysis for oxidation of volatile organic compounds: Part 2. Ozone decomposition and deactivation of γ-Al2O3. *Applied catalysis B: environmental* **2005,** *58* (3-4), 217-226.
30. Holzer, F.; Kopinke, F.; Roland, U., Influence of ferroelectric materials and catalysts on the performance of non-thermal plasma (NTP) for the removal of air pollutants. *Plasma Chemistry and Plasma Processing* **2005,** *25* (6), 595-611.
31. Hensel, K.; Katsura, S.; Mizuno, A., DC microdischarges inside porous ceramics. *IEEE Transactions on Plasma Science* **2005,** *33* (2), 574-575.
32. Hensel, K.; Martišovitš, V.; Machala, Z.; Janda, M.; Leštinský, M.; Tardiveau, P.; Mizuno, A., Electrical and optical properties of AC microdischarges in porous ceramics. *Plasma Processes and Polymers* **2007,** *4* (7-8), 682-693.
33. Suttikul, T.; Sreethawong, T.; Sekiguchi, H.; Chavadej., S., Ethylene epoxidation over alumina-and silica-supported silver catalysts in low-temperature ac dielectric barrier discharge. *Plasma Chemistry and Plasma Processing* **2011,** *31* (2), 273-290.
34. Wannagat, U. *The Silicon-Nitrogen Bond*; Springer, Boston, MA,, 1978; pp 77-90.
35. James, B. D.; DeSantis, D. A.; Saur, G. *Hydrogen Production Pathways Cost Analysis (2013–2016)*; Strategic Analysis Inc., Arlington, VA (United States): 2016.
36. Taguchi, A.; Schüth., F., Ordered mesoporous materials in catalysis. *Microporous and mesoporous materials* **2005,** *77* (1), 1-45.
37. Rao, Y.; Antonelli., D. M., Mesoporous transition metal oxides: characterization and applications in heterogeneous catalysis. *Journal of Materials Chemistry* **2009,** *19* (14), 1937-1944.

38. Beck, J. S., A new family of mesoporous molecular sieves prepared with liquid crystal templates. . *Journal of the American Chemical Society* **1992,** *114*, 10834-10843.
39. Kresge, C. T.; Leonowicz, M. E.; Roth, W. J.; Vartuli, J. C.; Beck, J. S., Ordered mesoporous molecular sieves synthesized by a liquid-crystal template mechanism. *Nature* **1992,** *359*, 710-712.
40. Carreon, M. A.; Guliants., V. V., Ordered meso-and macroporous binary and mixed metal oxides. . *European journal of inorganic chemistry* **2005,** *2005*, 27-43. .
41. Shah, J.; Wu, T.; Lucero, J.; Carreon, M. A.; Carreon, M. L., Nonthermal Plasma Synthesis of Ammonia over Ni-MOF-74. *ACS Sustainable Chemistry & Engineering* **2018,** *7* (1), 377-383.
42. Shah, J. R.; Gorky, F.; Lucero, J.; Carreon, M. A.; Carreon, M. L., Ammonia Synthesis via Atmospheric Plasma Catalysis: Zeolite 5A, a Case of Study. *Ind. Eng. Chem. Res.* **2020,** *59* (11), 5167-5176.
43. Gorky, F.; Carreon, M. A.; Carreon, M. L., Experimental Strategies to Increase Ammonia Yield in Plasma Catalysis over LTA and BEA Zeolites. *IOP SciNotes* **2020,** *1* (2), 024801.
44. Gorky, F.; Guthrie, S. R.; Smoljan, C. S.; Crawford, J. M.; Carreon, M. A.; Carreon, M. L., Plasma ammonia synthesis over mesoporous silica SBA-15. *J. Phys. D* **2021,** *54* (26), 264003.
45. van't Veer, K.; Reniers, F.; Bogaerts, A., Zero-Dimensional Modeling of Unpacked and Packed Bed Dielectric Barrier Discharges: The Role of Vibrational Kinetics in Ammonia Synthesis. *Plasma Sources Sci Technol* **2020,** *29* (4), 045020.
46. Brandenburg, R., Dielectric barrier discharges: progress on plasma sources and on the understanding of regimes and single filaments. *Plasma Sources Science and Technology* **2017,** *26* (5), 053001.
47. Gao, M.; Zhang, Y.; Wang, H.; Guo, B.; Zhang, Q.; Bogaerts, A., Mode transition of filaments in packed-bed dielectric barrier discharges. *Catalysts* **2018,** *8* (6), 248.
48. Hala, A., Plasma dynamics in a packed bed dielectric barrier discharge (DBD) operated in helium. *Journal of Physics D: Applied Physics* **2018,** *51* (11), 11LT02.
49. Kruszelnicki, J.; Hala, A.; Kushner, M. J., Formation of surface ionization waves in a plasma enhanced packed bed reactor for catalysis applications. *Chemical Engineering Journal* **2020,** *382*, 123038.
50. Engeling, K. W.; Kruszelnicki, J.; Kushner, M. J.; Foster, J. E., Time-resolved evolution of micro-discharges, surface ionization waves and plasma propagation in a two-dimensional packed bed reactor. *Plasma Sources Science and Technology* **2018,** *27* (8), 085002.
51. Cheng, H.; Ma, M.; Zhang, Y.; Liu, D.; Lu, X., The plasma enhanced surface reactions in a packed bed dielectric barrier discharge reactor. *Journal of Physics D: Applied Physics* **2020,** *53* (14), 144001.
52. Li, Y.; Yang, D.-Z.; Qiao, J.-J.; Zhang, L.; Wang, W.-Z.; Zhao, Z.-L.; Zhou, X.-F.; Yuan, H.; Wang, W.-C., The dynamic evolution and interaction with dielectric material of the discharge in packed bed reactor. *Plasma Sources Science and Technology* **2020,** *29* (5), 055004.
53. Shah, J.; Wang, W.; Bogaerts, A.; Carreon, M. L., Ammonia Synthesis by Radio Frequency Plasma Catalysis: Revealing the Underlying Mechanisms. *ACS Appl. Energy Mater.* **2018,** *1* (9), 4824-4839.
54. Dongyuan, Z.; Sun, J.; Li, Q.; Stucky, G. D., Morphological control of highly ordered mesoporous silica SBA-15. *Chemistry of materials* **2000,** *12*, 275-279.
55. Crawford, J. M.; Anderson, R.; Gasvoda, R. J.; Kovach, N. C.; Smoljan, C. S.; Jasinski, J. B.; Trewyn, B. G.; Agarwal, S.; Gómez-Gualdrón, D. A.; Carreon, M. A., Vacancy Healing as a Desorption Tool: Oxygen Triggered Removal of Stored Ammonia from NiO1–x/MOR Validated by Experiments and Simulations. *ACS Applied Energy Materials* **2020,** *3* (9), 8233-8239.
56. Gorky, F.; Lucero, J. M.; Crawford, J. M.; Blake, B.; Carreon, M. A.; Carreon, M. L., Plasma-Induced Catalytic Conversion of Nitrogen and Hydrogen to Ammonia over Zeolitic Imidazolate Frameworks ZIF-8 and ZIF-67. *ACS Appl. Mater. Interfaces* **2021,** *13* (18), 21338-21348.
57. Gorky, F.; Best, A.; Jasinski, J.; Allen, B. J.; Alba-Rubio, A. C.; Carreon, M. L., Plasma catalytic ammonia synthesis on Ni nanoparticles: The size effect. *J. Catal.* **2021,** *393*, 369-380.
58. Hong, T.; De León, F., Lissajous curve methods for the identification of nonlinear circuits: Calculation of a physical consistent reactive power. *IEEE Transactions on Circuits and Systems I: Regular Papers* **2015,** *62* (12), 2874-2885.
59. Peeters, F. J. J.; Van de Sanden, M. C. M., The influence of partial surface discharging on the electrical characterization of DBDs. *Plasma Sources Science and Technology* **2014,** *24* (1), 015016.
60. Gordon, I. E.; Rothman, L. S.; Hargreaves, R. J.; Hashemi, R.; Karlovets, E. V.; Skinner, F. M.; Conway, E. K.; Hill, C.; Kochanov, R. V.; Tan, Y., The HITRAN2020 molecular spectroscopic database. *Journal of quantitative spectroscopy and radiative transfer* **2021**, 107949.
61. C.O., L. In *Optical Diagnostics and Collisional-Radiative Models*, VKI Course on Hypersonic Entry and Cruise Vehicles, Stanford University, Stanford University, June 30 – July 3, 2008.
62. Pearse, R. W. B.; Gaydon, A. G., *The identification of molecular spectra*. Chapman and Hall London: 1976; Vol. 297.
63. Belostotskiy, S. G.; Ouk, T.; Donnelly, V. M.; Economou, D. J.; Sadeghi, N., Gas temperature and electron density profiles in an argon dc microdischarge measured by optical emission spectroscopy. *Journal of applied physics* **2010,** *107* (5), 053305.
64. Mijatović, Z.; Djurović, S.; Gavanski, L.; Gajo, T.; Favre, A.; Morel, V.; Bultel, A., Plasma density determination by using hydrogen Balmer Hα spectral line with improved accuracy. *Spectrochimica Acta Part B: Atomic Spectroscopy* **2020,** *166*, 105821.
65. Gigosos, M. A.; Gonzalez, M. A.; Cardenoso, V., Computer simulated Balmer-alpha,-beta and-gamma Stark line profiles for non-equilibrium plasmas diagnostics. *Spectrochimica Acta Part B: Atomic Spectroscopy* **2003,** *58* (8), 1489-1504.
66. Gigosos, M. A.; Cardeñoso, V., New plasma diagnosis tables of hydrogen Stark broadening including ion dynamics. *Journal of Physics B: Atomic, Molecular and Optical Physics* **1996,** *29* (20), 4795.
67. Van der Horst, R. M.; Verreycken, T.; Van Veldhuizen, E. M.; Bruggeman, P. J., Time-resolved optical emission spectroscopy of nanosecond pulsed discharges in atmospheric-pressure N2 and N2/H2O mixtures. *Journal of Physics D: Applied Physics* **2012,** *45* (34), 345201.
68. Kepple, P.; Griem, H. R., Improved Stark profile calculations for the hydrogen lines H α, H β, H γ, and H δ. *Physical Review* **1968,** *173* (1), 317.
69. Griem, H. R., Plasma Spectroscopy, Mc Graw-Hill, New York (1964). *Spectral Line Broadening by Plasmas* **1966**.
70. Drake, G. W. F., *Atomic, Molecular and Optical Physics*. Springer New York: 2006.

71. Patil, B. S.; Cherkasov, N.; Srinath, N. V.; Lang, J.; Ibhadon, A. O.; Wang, Q.; Hessel, V., The role of heterogeneous catalysts in the plasma-catalytic ammonia synthesis. *Catalysis Today* **2021,** *362*, 2-10.
72. Barboun, P.; Mehta, P.; Herrera, F. A.; Go, D. B.; Schneider, W. F.; Hicks, J. C., Distinguishing Plasma Contributions to Catalyst Performance in Plasma-Assisted Ammonia Synthesis. *ACS Sustain. Chem. Eng.* **2019,** *7* (9), 8621-8630.
73. Pipa, A. V.; Brandenburg, R., The equivalent circuit approach for the electrical diagnostics of dielectric barrier discharges: The classical theory and recent developments. *Atoms* **2019,** *7* (1), 14.
74. Baskaran, S.; Liu, J.; Domansky, K.; Kohler, N.; Li, X.; Coyle, C.; Fryxell, G. E.; Thevuthasan, S.; Williford, R. E., Low dielectric constant mesoporous silica films through molecularly templated synthesis. *Advanced Materials* **2000,** *12* (4), 291-294.
75. Wang, Y.; Craven, M.; Yu, X.; Ding, J.; Bryant, P.; Huang, J.; Tu, X., Plasma-Enhanced Catalytic Synthesis of Ammonia over a Ni/Al2O3 Catalyst at Near-Room Temperature: Insights into the Importance of the Catalyst Surface on the Reaction Mechanism. *ACS Catal.* **2019,** *9* (12), 10780-10793.
76. Lu, X.; Naidis, G. V.; Laroussi, M.; Reuter, S.; Graves, D. B.; Ostrikov, K., Reactive species in non-equilibrium atmospheric-pressure plasmas: Generation, transport, and biological effects. *Physics Reports* **2016,** *630*, 1-84.
77. Neyts, E. C.; Ostrikov, K.; Sunkara, M. K.; Bogaerts, A., Plasma catalysis: synergistic effects at the nanoscale. *Chemical reviews* **2015,** *115* (24), 13408-13446.
78. Liu, T.-W.; Gorky, F.; Carreon, M.; Gualdron, D. A., Energetics of pathways enabled by experimentally detected radicals during catalytic, plasma-assisted NH3 synthesis. **2021**.
79. Mehta, P.; Barboun, P.; Herrera, F. A.; Kim, J.; Rumbach, P.; Go, D. B.; Hicks, J. C.; Schneider, W. F., Overcoming Ammonia Synthesis Scaling Relations with Plasma-Enabled Catalysis. **2018,** *1* (4), 269-275.

**TOC (Table of Contents/ Abstract Graphic)**

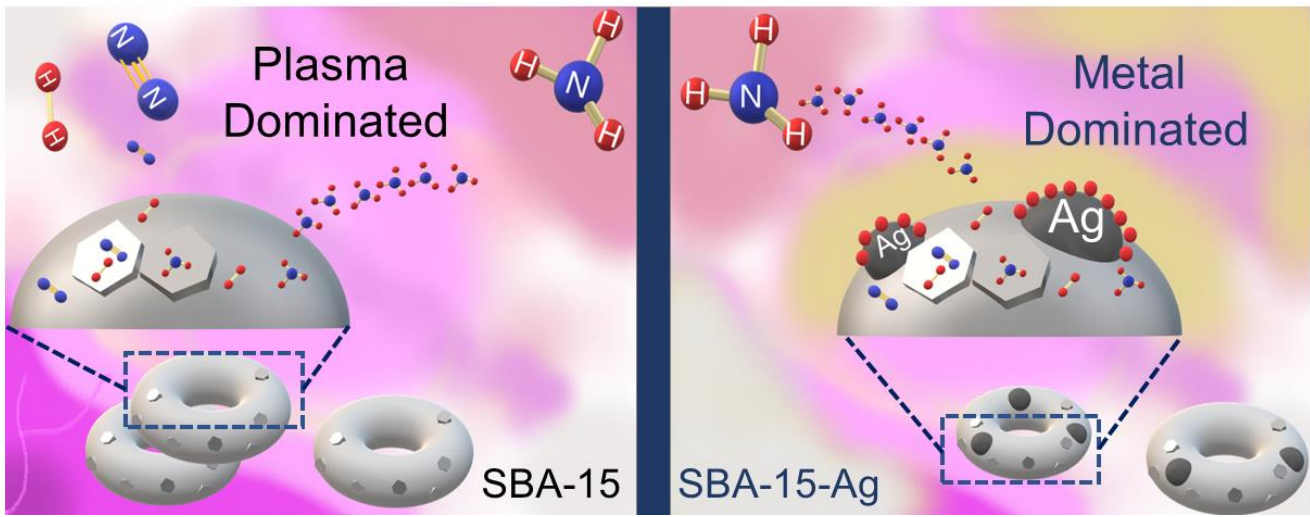

**TOC.** Mesoporous materials for Plasma Catalytic Ammonia Synthesis, leading to different regimes where surface and metal dominated regimes lead to the production of ammonia respectively.